\documentstyle[preprint,aps]{revtex}
 \begin{document}
\title{Remarks on Gauge-Invariant Variables and Interaction Energy in QED}
\author{ P. Gaete$^1$\thanks{E-mail: pgaete@fis.utfsm.cl}} 
\address{Departamento de F\'{i}sica, Universidad T\'{e}cnica Federico Santa\\
Maria, Casilla110-V, Valparaiso, Chile} 
\maketitle
\begin{abstract}
The calculation of the interaction energy in pure QED and
Maxwell-Chern-Simons gauge theory is re-examined by exploiting the path
dependence of the gauge-invariant variables formalism. In particular, we
consider a spacelike straight line which leads to the Poincar\'{e} gauge.
Subtleties related to the problem of exhibiting explicitly the interaction
energies are illustrated.
\end{abstract}
\section{INTRODUCTION}

It is well known that the development of an analytical understanding of
non-perturbative aspects of quantum chromodynamics (QCD) is a subject of
intense study. As a step towards this goal there is a renewed recent
interest in formulations of QCD in which gauge invariant variables are
explicitly constructed\cite{Lavelle,Prokhorov} . For instance, quarks are
considered as dressed objects, where this dressing can be viewed as
surrounding the quark with a cloud of gauge fields; all this in connection
with the confinement problem.

On the other hand, this last remark about the gauge-invariant or dressing
field is no new. It was Dirac\cite{Dirac} who originally proposed to use
for the electron the gauge- invariant field

\begin{equation}
\Psi =\exp \left( ie\text{ }\frac{\partial _{i}A^{i}}{\mathbf{\nabla }^{%
\mathbf{2}}}\right) \psi ,  \label{expa}
\end{equation}
as a physical variable. Consequently, $\Psi $ describes creation and
annihilation of charges together with their proper electric field, where
this field gives rise to the static interaction that obeys Coulomb$^{\text{'}%
}$s law. It is worthwhile remarking at this point that Dirac's expression
for the electron field is both non-local and non-covariant. Hence the
physical electron in QED is dressed, and as we mentioned before, this
dressing looks like a cloud of gauge fields surrounding the charged
particles with the characteristic feature that this cloud spreads out over
the whole space, giving rise to a non-local object. Furthermore, we
recognize that Dirac's electron field is written in the Coulomb gauge,
whence one can appreciate the link between the physical or dressed variables
and gauge fixing. Recently, by using a gauge-invariant but path dependent
formalism in abelian gauge theories, we illustrated how the gauge fixing
procedure corresponds, in this formalism, to a path choice. We therefore
developed a path-dependent but physical QED where a consistent quantization
directly in the path space was carried out\cite{Gaete}.

Thus, our concern in this Brief Report is to reconsider the calculation of
the interaction energy between point-like sources in QED, paying due
attention to the structure of the fields that surround the charges by
exploiting the path dependence of the gauge-invariant variables formalism.
In Sec.II we reexamine this calculation in pure QED or, more precisely,
electrodynamics of massive charges, and in Maxwell-Chern-Simons gauge
theories.

\section{INTERACTION \ ENERGY}

\subsection{PURE QED}

In order to calculate the energy of the external fields of static charges,
we shall begin by recalling the expression for a physical electron\cite{Dirac,Gaete},

\begin{equation}
\Psi (y)=\exp \left( -ie\int_{C_{\xi y}}dz^{\mu }A_{\mu }(z)\right) \psi (y),
\label{expb}
\end{equation}
where the integral is taken along some path or contour $C_{\xi y\text{ }}$%
connecting $\xi $ and $y$. As already expressed, the main point is that we
can choose a particular gauge condition by selecting a specific path or
contour. Thus we now choose the path as a spacelike straight line which
leads to the Poincar\'{e} gauge \cite{Gaete}: $z^{k}=$ $\xi ^{k}$ + $%
\alpha (y-\xi )^{k}$, where $\alpha $ ($0\le \alpha \le 1$) 
is the parameter describing the contour and $\xi $ is a fixed point
(reference point). There is no essential loss of generality if we restrict
our considerations to $\xi ^{k}=0.$ Accordingly, in the Poincar\'{e} gauge, (%
\ref{expb}) becomes 
\begin{equation}
\Psi (y)=\exp \left( -ie\int_{0}^{y}\text{ }dz^{k}A_{k}(z)\right) \psi (y)
\label{expc}
\end{equation}
Two remarks are pertinent at this point. First, Eq.(\ref{expc}) does not
follow from Eq.(\ref{expb}) in the temporal gauge. It was shown in Ref. \cite{Gaete} that to attain the temporal gauge we have to choose a timelike
straight line: $z^{0}=t^{0}+\alpha $ $(x^{0}-t^{0})$, where $t^{0}$ is a
arbitrary parameter. Second, it should be noted that with respect to gauge
transformations the physical electron (\ref{expc}) acquire a phase factor $%
\exp \left( -ie\Lambda (\xi =0)\right) $; as well as, since the point $\xi =0
$ is assumed to be fixed, the translational invariance of the physical
electron is broken. These drawbacks are avoided by letting to point $\xi $
go to infinity.

At the same time (\ref{expc}) represents charged particles with a static
electric field on a line or, more precisely, on a tube. To see this more
clearly let $\mid $ $E\rangle $ be an eigenvector of the electric field
operator $E_{i}(x)$, with eigenvalue $\varepsilon _{i}(x)$ :

\begin{equation}
E_{i}(x)\mid E\rangle =\varepsilon _{i}(x)\mid E\rangle .  \label{eiga}
\end{equation}
Next we will consider the state $\Psi (y)\mid E\rangle $. Using (\ref{eiga})
we obtain

\begin{equation}
E_{i}(x)\text{ }\Psi (y)\mid E\rangle =\Psi (y)\text{ }E_{i}(x)\mid E\rangle
+\left[ E_{i}(x),\Psi (y)\right] \mid E\rangle .  \label{eigb}
\end{equation}
Using the Hamiltonian formalism developed in Ref. \cite{Gaete}, we find that

\begin{equation}
E_{i}(x)\text{ }\Psi (y)\mid E\rangle =\left( \varepsilon
_{i}(x)+e\int_{0}^{1}d\alpha \text{ }y_{i}\text{ }\delta ^{(3)}\left( \alpha 
\mathbf{y}-\mathbf{x}\right) \right) \Psi (y)\mid E\rangle .  \label{eigc}
\end{equation}
We have thus verified that the operator $\Psi (y)$ is the dressing operator
of creation of an electron together with their proper electric field (or the
operator of absorption of a positron). Moreover, the above result clearly
proves that we have a static electric field on a line (tube), because the
integral in (\ref{eigc}) is nonvanishing only on the contour of integration.
It is perhaps worth mentioning at this stage that if we consider a modified
form for the physical electron in the Poincar\'{e} gauge (\ref{expc}), which
is equivalent to the Coulomb gauge \cite{Gaete}, that is, 
\begin{equation}
\Psi (y)=\exp \left( -ie\int_{0}^{y}dz^{k}A_{k}^{L}(z)\right) \psi (y),
\label{expd}
\end{equation}
where $A_{k}^{L}$ refers to the longitudinal part of $A_{k}$, we would
obtain that the field $\Psi $ dresses the charge $\psi $ with the static
Coulomb electric field, in other words, 
\begin{equation}
E_{i}(x)\text{ }\Psi (y)\mid E\rangle =\left( \varepsilon _{i}(x)+\frac{e}{%
4\pi }\frac{\mathbf{x}_{i}-\mathbf{y}_{i}}{\mid \mathbf{x-y\mid }^{3}}%
\right) \Psi (y)\mid E\rangle .  \label{elec}
\end{equation}
For more details about the comparison between the Poincar\'{e} and Coulomb
gauges we refer to \cite{Galvao}.

Now we compute the energy of the external field of static charges, where a
fermion is localized at $\mathbf{y}^{\prime }$ and an antifermion at $%
\mathbf{y.}$ So we proceed to calculate the mean value of the
electromagnetic energy operator $H$ in the physical state $\mid \Omega
\rangle $ , which we will denote by $\langle H\rangle _{\Omega }.$ From the
Hamiltonian framework presented in Ref. \cite{Gaete}, $\langle H\rangle
_{\Omega }$ is given by 
\begin{equation}
\langle H\rangle _{\Omega }=\langle \Omega \mid \int d^{3}x\text{ }\left( -%
\frac{1}{2}\pi _{i}\pi ^{i}+\frac{1}{4}F_{ij}F^{ij}\right) \mid \Omega
\rangle .  \label{enea}
\end{equation}
As mentioned before, the fermions are taken to be infinitely massive
(static), which means that there is no magnetic field. In such a case the
expression (\ref{enea}) then becomes 
\begin{equation}
\langle H\rangle _{\Omega }=\langle \Omega \mid \frac{1}{2}\int d^{3}x\text{ 
}\mathbf{E}^{\mathbf{2}}(x)\mid \Omega \rangle .  \label{eneb}
\end{equation}
As it has been established by Dirac \cite{Dirac}, the physical states $\mid
\Omega \rangle $ are characterized by the gauge invariant ones. Thus a state
which has a fermion at $\mathbf{y}^{\prime }$ and an antifermion at $\mathbf{%
y}$ is given by 
\begin{equation}
\mid \Omega \rangle \equiv \mid \overline{\Psi }(\mathbf{y})\Psi (\mathbf{y}%
^{\prime })\rangle =\overline{\psi }(\mathbf{y})\exp \left( -ie\int_{\mathbf{%
y}}^{\mathbf{y}^{\prime }}dz^{i}A_{i}(z)\right) \psi (\mathbf{y}^{\prime
})\mid 0\rangle ,  \label{inva}
\end{equation}
where $\mid 0\rangle $ is the physical vacuum state. Following steps similar
to those leading to Eq.(\ref{eigc}) we obtain 
\begin{equation}
E_{i}(x)\mid \overline{\Psi }(\mathbf{y})\Psi (\mathbf{y}^{\prime })\rangle =%
\overline{\Psi }(\mathbf{y})\Psi (\mathbf{y}^{\prime })E_{i}(x)\mid 0\rangle
+e\int_{\mathbf{y}}^{\mathbf{y}^{\prime }}dz_{i}\delta ^{(3)}(\mathbf{x}-%
\mathbf{z})\mid \overline{\Psi }(\mathbf{y})\Psi (\mathbf{y}^{\prime
})\rangle  \label{invb}
\end{equation}
$\qquad $\linebreak By using (\ref{invb}) we then evaluate the energy in
presence of the static charges, yielding 
\begin{equation}
\langle H\rangle _{\Omega }=\langle H\rangle _{0}+\frac{e^{2}}{2}\int_{%
\mathbf{y}}^{\mathbf{y}^{\prime }}dz^{i}\int_{\mathbf{y}}^{\mathbf{y}%
^{\prime }}dz_{i}^{\prime }\text{ }\delta ^{(3)}(\mathbf{z}-\mathbf{z}%
^{\prime }),  \label{enec}
\end{equation}
where $\langle H\rangle _{0}=\langle 0\mid H\mid 0\rangle .$ We emphasize
that the second term in the r.h.s. of (\ref{enec}) corresponds to an
interaction energy of the external fields of static charges, after
substracting the divergent energies associated with the single particle
states. Remembering that the integrals over $z_{i}$ and $z_{i}^{\prime }$
are zero except on the contour of integrations, one obtains the following
interaction energy 
\begin{equation}
V=\frac{e^{2}}{2}\text{ }k\mid \mathbf{y}-\mathbf{y}^{\prime }\mid \text{ },
\label{pota}
\end{equation}
where $k=\delta ^{(2)}(0).$

Special care has to be exercised since the physical interpretation of (\ref
{pota}) is not clear in the literature and its discussion must be amended \cite{Axial}. For instance, it has been argued \cite{Fursaev} that the
energy (\ref{pota}) is unstable, that is, it breaks down into
electromagnetic radiation and the Coulomb field of two opposite charges.
Such a picture, in our opinion, is certainly debatable because the sources
of the external fields are stationary. In fact, we now show that although
the Coulomb interaction does not appear explicitly in the quantity $\frac{%
e^{2}}{2}\int d^{3}x\left( \int_{\mathbf{y}}^{\mathbf{y\prime }}dz_{i}\delta
^{(3)}(\mathbf{x-z})\right) ^{2}$, this expression is nothing but the
Coulomb interaction plus an infinite self-energy term. For this, we focus
our attention to 
\begin{equation}
V=\frac{e^{2}}{2}\int d^{3}x\text{ }\left( \int_{\mathbf{y}}^{\mathbf{y}%
^{\prime }}\text{ }dz_{i}\delta ^{(3)}(\mathbf{x}-\mathbf{z})\right) ^{2}.
\label{potb}
\end{equation}
As already expressed, in order to carry out this calculation we write the
path as $\mathbf{z=y+}\alpha \mathbf{(y}^{\prime }-\mathbf{y)}$, where $%
\alpha $ is the parameter describing the contour. In this case we have that
\begin{equation}
\int_{\mathbf{y}}^{y\prime }dz_{i}\delta ^{(3)}\left( \mathbf{x-z}\right) =(%
\mathbf{y\prime -y)}\int_{0}^{1}d\alpha \delta ^{(3)}\left( \mathbf{y-x}%
\text{ }+\alpha \left( \mathbf{y}^{\prime }-\mathbf{y}\right) \right) .
\label{dela}
\end{equation}
This can be conveniently written using spherical coordinates as 
\begin{equation}
\int_{\mathbf{y}}^{\mathbf{y\prime }}dz_{i}\delta ^{(3)}\left( \mathbf{x-z}%
\right) =\frac{(\mathbf{y-y}^{\prime })}{\mid \mathbf{y-y}^{\prime }\mid ^{2}%
}\int_{0}^{1}d\alpha \frac{1}{\alpha ^{2}}\delta \left( \mid \mathbf{y-x\mid
-}\alpha \mathbf{\mid y}^{\prime }-\mathbf{y}\mid \right)
\sum_{l,m}Y_{lm}^{*}(\theta ^{\prime },\phi ^{\prime })Y_{lm}(\theta ,\phi ),
\label{delb}
\end{equation}
hence expression (\ref{delb}) reduces to
\begin{equation}
\int_{\mathbf{y}}^{\mathbf{y\prime }}dz_{i}\delta ^{(3)}\left( \mathbf{x-z}%
\right) =\frac{(\mathbf{y}^{\prime }-\mathbf{y)}}{\mid \mathbf{y}^{\prime }-%
\mathbf{y\mid }}\frac{1}{\mid \mathbf{y-x\mid }^{2}}\sum_{l,m}Y_{lm}^{*}(%
\theta ^{\prime },\phi ^{\prime })Y_{lm}(\theta ,\phi ).  \label{delc}
\end{equation}
According to (\ref{delc}), the expression for the interaction energy (\ref
{potb}) reads  
\begin{equation}
V=\frac{e^{2}}{2}\int d^{3}x\left( \frac{\mathbf{y}^{\prime }-\mathbf{y}}{%
\mid \mathbf{y}^{\prime }-\mathbf{y}\mid }\text{ }\frac{1}{\mid \mathbf{y}-%
\mathbf{x}\mid ^{2}}\sum_{l,m}Y_{lm}^{*}(\theta ^{\prime },\phi ^{\prime
})Y_{lm}(\theta ,\phi )\right) ^{2}.  \label{sat}
\end{equation}
Introducing the integration variable $\mathbf{r=y-x}$, and using usual
properties for the spherical harmonics, we obtain
\begin{equation}
V=\frac{e^{2}}{2}\int_{0}^{\vert {\bf y} - {\bf y^{'}} \vert} \frac{dr}{r^2}, 
\label{deld}
\end{equation}
with $r=\vert {\bf r}\vert$. In other words, we find that

\begin{equation}
V=-\frac{e^{2}}{4\pi }\frac{1}{\mid \mathbf{y}^{\prime }-\mathbf{y}\mid }
\label{sata}
\end{equation}
after substracting the self-energy term.

To end this subsection we also draw attention to the fact that with the path
choice stated in (\ref{expc})(modified Poincar\'{e} gauge) which is
equivalent to the Coulomb gauge, and from the formalism developed in 
\cite{Gaete}, we can write a scalar potential as 
\begin{equation}
\mathcal{A}_{0}(t,\mathbf{x})=-\int_{0}^{1}d\alpha \text{ }\mathbf{x}\cdot 
\mathbf{E}^{L}(t,\alpha \mathbf{x}),  \label{esca}
\end{equation}
which, by employing the Hamiltonian structure of pure QED, may be rewritten
as 
\begin{equation}
\mathcal{A}_{0}(t,\mathbf{x})=\int_{0}^{1}d\alpha \text{ }%
x^{i}E_{i}^{L}(t,\alpha \mathbf{x})=-\int_{0}^{1}d\alpha \text{ }\frac{%
x^{i}\partial _{i}^{\alpha \mathbf{x}}}{\mathbf{\nabla }_{\mathbf{\alpha x}%
}^{\mathbf{2}}}J^{0}(\alpha \mathbf{x}),  \label{escb}
\end{equation}
where the superscript $L$ refers to the longitudinal part and $J^{0}$ is the
external source. As a consequence of this, the static potential $V$ for a
pair of static point-like opposite charges located at $\mathbf{y}$ and $%
\mathbf{y}^{\prime }$, that is, $J^{0}(t,\mathbf{x})=e\left( \delta ^{(3)}(%
\mathbf{x-y})-\delta ^{(3)}(\mathbf{x-y\prime )}\right) $, is given by 
\begin{equation}
V=\text{ }e\text{ }\left( \mathcal{A}_{0}(\mathbf{y})-\mathcal{A}_{0}(%
\mathbf{y\prime })\right) =-\frac{e^{2}}{4\pi }\text{ }2\text{ }\frac{1}{%
\mid \mathbf{y\prime }-\mathbf{y}\mid }\text{ ,}  \label{difa}
\end{equation}
after substracting the self-energy term. Let us point out that (\ref{difa})
is the same expression as in Ref. \cite{Haagensen}, which find here an
independent derivation.

\subsection{ MAXWELL-CHERN-SIMONS GAUGE THEORY}

We now consider the calculation of the interaction energy between static
point-like sources in a topologically massive gauge theory. In such a case
the Lagrangian reads \cite{Jackiw}: 
\begin{equation}
L=-\frac{1}{4}\text{ }F_{\mu \nu }^{2}+\frac{1}{4}\text{ }\theta \text{ }%
\varepsilon ^{\mu \nu \rho }A_{\mu }F_{\nu \rho }-A_{0}J^{0},  \label{topa}
\end{equation}
where $J^{0}$ is the external current and $\theta $ is the topological mass.

Before we proceed to work out explicitly the energy, we shall begin by
summarizing the canonical quantization of the theory (\ref{topa}) from the
Hamiltonian analysis point of view. The canonical momenta are $\pi ^{\mu
}=-F^{0\mu }+\frac{\theta }{2}\varepsilon ^{0\mu \nu }A_{\nu }$ with the
only nonvanishing canonical Poisson brackets being 
\begin{equation}
\text{ }\{\text{ }A_{\mu }(t,\mathbf{x),\pi }^{\nu }(t,\mathbf{y)\}=\delta }%
_{\mu }^{\nu }\text{ }\delta ^{(2)}(\mathbf{x-y).}  \label{poi}
\end{equation}
As we can see there is one primary constraint, $\pi ^{0}=0$, and $\pi
^{i}=F^{i0}+\frac{\theta }{2}\varepsilon ^{ij}A_{j}$ $(i,j=1,2).$ So the
canonical Hamiltonian is 
\begin{equation}
H_{c}=\int d^{2}x\left( -\frac{1}{2}F_{i0}F^{i0}+\frac{1}{4}F^{ij}F_{ij}+\pi
_{i}\partial ^{i}A_{0}-\frac{\theta }{2}\varepsilon ^{ij}A_{0}\partial
_{i}A_{j}+A_{0}J^{0}\right) .  \label{cana}
\end{equation}
The conservation in time of the constraint $\pi ^{0}$ leads to the secondary
constraint (Gauss law): 
\begin{equation}
\Omega _{1}\left( x\right) =\partial _{i}\pi ^{i}+\frac{\theta }{2}%
\varepsilon _{ij}\partial ^{i}A^{j}-J^{0}=0.  \label{vina}
\end{equation}
There are no more constraints in the theory and the two we have found are
first class. The corresponding total (first class) Hamiltonian that
generates the time evolution of the dynamical variables is given by 
\begin{equation}
H=H_{c}+\int d^{2}x\text{ }\left( c_{0}(x)\pi _{0}(x)+c_{1}(x)\pi
_{1}(x)\right) ,  \label{hama}
\end{equation}
where $c_{0}(x)$ and $c_{1}(x)$ are arbitrary functions. Since $\pi ^{0}=0$
for all time and $\stackrel{\mathbf{\cdot }}{A_{0}}(x)=\left[
A_{0}(x),H\right] =c_{0}(x)$, which is arbitrary, we discard $A_{0}(x)$ and $%
\pi _{0}(x)$. In fact, it is redundant to retain the term containing $A_{0}$
because it can be absorbed by redefining the function $c_{1}(x)$. In this
case, (\ref{hama}) takes the form 
\begin{equation}
H=\int d^{2}x\left( -\frac{1}{2}F_{i0}F^{i0}+\frac{1}{4}F_{ij}F^{ij}+c^{%
\prime }(x)\left( \partial _{i}\pi ^{i}+\frac{\theta }{2}\varepsilon
_{ij}\partial ^{i}A^{j}-J^{0}\right) \right) ,  \label{hamb}
\end{equation}
where $c^{\prime }(x)=c_{1}(x)-A_{0}(x)$.

According to the standard procedure we impose one gauge constraint such that
the full set of constraints become second-class. Just as for the pure QED
case, we write the gauge fixing as follows 
\begin{equation}
\Omega _{2}(x)=\int_{0}^{1}d\alpha \text{ }x^{i}A_{i}(\alpha \mathbf{x)=0,}
\label{gfa}
\end{equation}
where $\alpha $ is the parameter describing a spacelike straight line of
integration. In this way, one easily verifies that the fundamental Dirac
brackets reads 
\begin{equation}
\left\{ A_{i}(x),A^{j}(y)\right\} ^{*}=0=\left\{ \pi _{i}(x),\pi
^{j}(y)\right\} ^{*},  \label{para}
\end{equation}
\begin{equation}
\left\{ A_{i}(x),\pi ^{j}(y)\right\} ^{*}=\delta _{i}^{j}\delta ^{(2)}\left( 
\mathbf{x-y}\right) -\partial _{i}^{x}\int_{0}^{1}d\alpha \text{ }%
x^{j}\delta ^{(2)}\left( \alpha \mathbf{x-y}\right) .  \label{parb}
\end{equation}
In order to illustrate the discussion, we now write the equations of motion
in terms of the magnetic $\left( B=\varepsilon _{ij}\partial
^{i}A^{j}\right) $ and electric $(E^{i}=\pi ^{i}-\frac{\theta }{2}%
\varepsilon ^{ij}A_{j})$ fields as 
\begin{equation}
\stackrel{\cdot }{E_{i}}(x)=-2\theta \varepsilon _{ij}E^{j}(x)-\varepsilon
_{ij}\partial ^{j}B,  \label{equa}
\end{equation}
\begin{equation}
\stackrel{\mathbf{\cdot }}{B}(x)=-\text{ }\varepsilon _{ij}\partial
^{i}E^{j}.  \label{equb}
\end{equation}
In the same way, we write the Gauss law as 
\begin{equation}
\partial _{i}E_{L}^{i}+\theta B-J^{0}=0,  \label{gaus}
\end{equation}
where $E_{L}^{i}$ refers to the longitudinal part of $E^{i}.$

The following remark deserves to be mentioned. As in the preceding
subsection, we will compute the interaction energy between point-like
sources in the static approximation ( the limit of large fermion masses).
Accordingly, from the equations of motion, as opposed to the pure QED case,
we have static electromagnetic fields: 
\begin{equation}
B=-\text{ }\theta \text{ }\frac{J^{0}}{\mathbf{\nabla }^{2}-\theta ^{2}}%
\text{ },  \label{mag}
\end{equation}
\begin{equation}
E_{i}(x)=\frac{1}{\theta }\text{ }\partial _{i}B,\text{ }  \label{ele}
\end{equation}
where $\mathbf{\nabla }^{2}$ is the two-dimensional laplacian. For $J^{0}(t,%
\mathbf{x)=}$ $e\delta ^{(2)}(\mathbf{x-y),}$ expressions (\ref{mag}) and (%
\ref{ele}) immediately show that 
\begin{equation}
B(x)=\frac{e\theta }{2\pi }\text{ }K_{0}\left( \theta \mid \mathbf{x-y\mid }%
\right) ,  \label{kce}
\end{equation}
\begin{equation}
E^{i}(x)=-\text{ }\frac{e\theta }{2\pi }\text{ }\frac{\left( \mathbf{x-y}%
\right) ^{i}}{\mid \mathbf{x-y\mid }}\text{ }K_{1}\left( \theta \mid \mathbf{%
x-y\mid }\right) ,  \label{kun}
\end{equation}
where $K_{0}$ and $K_{1}$ are modified Bessel's functions. Their limiting
forms for small and large $\theta \mid \mathbf{x-y}\mid $ are: 
$E^{i}\sim \frac{1}{\mid \mathbf{x-y}\mid }$ and 
$B\sim \ln (\mid \mathbf{x-y}\mid )$ for $\mid \mathbf{x-y}\mid $ $\ll \frac{1}{\theta }$,
while they fall of exponentially $\sim $ 
$\exp (-\theta \mid \mathbf{x-y}\mid )$ for $\mid \mathbf{x-y}\mid $ $\gg \frac{1}{\theta }$. This shows
that the electromagnetic fields,while falling off exponentially for $\mid 
\mathbf{x-y}\mid $ $\gg \frac{1}{\theta }$, are localized near the charge.
It is worth mentioning that the above expressions were previously derived in
Ref. \cite{Shizuya} considering the anyon statistics and its variation with
wavelength. In that reference it was remarked that an anyon can be viewed as
a point charge surrounded by a gauge field cloud of size $\frac{1}{\theta }$. Let us also point out that the fields (\ref{kce}) and (\ref{kun}) are
present at the classical level.

For the sake of simplicity, we will now compute the interaction energy via
expression (\ref{esca}). Thus we have that 
\begin{equation}
\mathcal{A}_{0}(t,\mathbf{x)=}\int_{0}^{1}d\alpha \text{ }%
x^{i}E_{i}^{L}(t,\alpha \mathbf{x)=-}\int_{0}^{1}d\alpha \text{ }\frac{%
x^{i}\partial _{i}^{\alpha \mathbf{x}}}{\mathbf{\nabla }_{\alpha \mathbf{x}%
}^{2}-\theta ^{2}}J^{0}(\alpha \mathbf{x).}  \label{bea}
\end{equation}
For $J^{0}(t,\mathbf{x)=}$ $e\delta ^{(2)}(\mathbf{x-a)}$ expression (\ref
{bea}) then becomes 
\begin{equation}
\mathcal{A}_{0}(t,\mathbf{x)=}\frac{e}{2\pi }\left( K_{0}(\theta \mid 
\mathbf{x-a}\mid )-K_{0}(\theta \mid \mathbf{a}\mid )\right) .  \label{beb}
\end{equation}
By means of (\ref{bea}) we evaluate the interaction energy for a pair of
static point-like opposite charges at $\mathbf{y}$ and $\mathbf{y}^{\prime }$%
, as 
\begin{equation}
V=e\left( \mathcal{A}_{0}(\mathbf{y)-}\mathcal{A}_{0}(\mathbf{y}^{\prime
})\right) =-\frac{e^{2}}{\pi }K_{0}(\theta \mid \mathbf{y-y}^{\prime }\mid ),
\label{bec}
\end{equation}
after substracting the constant terms.

Finally, one comment is pertinent in this context. Recently \cite{Abdalla}, by using bosonization methods, it has been derived the interparticle
energy for three-dimensional massive quantum electrodynamics. By starting
from the three-dimensional massive QED in the covariant gauge and in the
presence of an external source $J^{0}$, these authors \cite{Abdalla}
obtain the bosonized version in the large mass limit (quadratic
approximation), which is the Maxwell-Chern-Simons theory in the covariant
gauge. Next, by using the Feynman propagator in two dimensions, they arrive
at the result (\ref{bec}) in the limit for heavy fermions $(m\rightarrow
\infty )$. However, the essential difference between our analysis and that
of Ref. \cite{Abdalla} lies on the observation that the potential (\ref{bec}%
) is present at classical level too.

\section{ACKNOWLEDGMENTS}

It is a pleasure to thank J. Gamboa and J. Zanelli for useful discussions
and strong encouragement. I would also like to thank I. Schmidt for his
support.


\begin{thebibliography}{}
\bibitem{Axial}  It should be emphasized that the same result is obtained
when we take an infinite refernce point in (\ref{expb}), as in Ref. 
\cite{Prokhorov}. In fact, it has been shown in Ref. \cite{Gaete} that if we
choose the path $C_{\xi x}$ in the form of a spacelike straight line, where
the reference point $\xi $ is in spatial infinity, such potentials are said
belong the axial gauge.
\end{thebibliography}
\end{document}